\documentclass[12pt]{nature} 
 
\begin{document} 
 
\title{Steric engineering of metal-halide perovskites\\ with tunable optical band gaps} 
 
\author{Marina R. Filip$^1$, Giles E. Eperon$^2$, Henry J. Snaith$^2$ \& Feliciano Giustino$^1$} 
 
\maketitle 

\vspace{0.2cm} 
\begin{affiliations} 
\item[$^1$] Department of Materials, University of Oxford, Parks Road, Oxford, OX1 3PH 
\vspace{-0.2cm} 
\item[$^2$] Department of Physics, University of Oxford, Clarendon Laboratory, Parks Road, Oxford, OX1 3PU, UK
\end{affiliations} 
 
\begin{abstract} 
Owing to their high energy-conversion efficiency and inexpensive fabrication routes, solar cells  
based on metal-organic halide perovskites have rapidly gained prominence as a disruptive technology.  
An attractive feature of perovskite absorbers is the possibility of tailoring their properties  
by changing the elemental composition through the chemical precursors. In this context, rational  
\textit{in silico} design represents a powerful tool for mapping the vast materials landscape and  
accelerating discovery. Here we show that the optical band gap of metal-halide perovskites, a key  
design parameter for solar cells, strongly correlates with a simple structural feature, 
the largest metal-halide-metal bond angle. Using this descriptor we suggest
continuous tunability of the optical gap from the mid-infrared to the visible. 
Precise band gap engineering is achieved by controlling the bond angles through the steric size 
of the molecular cation. Based on these design principles we predict novel low-gap perovskites 
for optimum photovoltaic efficiency, and we demonstrate the concept of band gap 
modulation by synthesising and characterising novel mixed-cation perovskites.

\end{abstract} 
 
\vspace{0.3cm} 
Hybrid organic-inorganic perovskites~\cite{Mitzi1994,Kagan1999,Miyasaka2009,Kanatzidis2012} recently surged to worldwide attention  
following the report of highly efficient solar cells based on methylammonium lead 
halides~\cite{Lee2012,Kim2012,Burschka2013,Liu2013}.  
In fact, within less than two years of development, the photovoltaic energy-conversion efficiency of 
these devices increased from the initial 10.9\%~\cite{Lee2012} to the current record 
of 17.9\%~(NREL).
This impressive trend fuelled the expectation that perovskite photovoltaics 
may soon revolutionise the solar energy landscape. 
 
Given the unusually long lifetimes of the photoexcited  
carriers~\cite{Stranks2013,Wehrenfennig2013,Xing2013} and the relatively low optical  
gap~\cite{Stoumpos2013,Noh2013}, it has been proposed that the perovskite layer may act  
both as the light absorber and as the electron and hole transporter~\cite{Snaith2013}, 
similarly to bulk semiconductor solar cells. Within this scenario, it is expected that the 
energy-conversion efficiency can be increased by following the standard Schockley-Queisser 
analysis, that is by reducing the optical band gap to values near 1.3-1.4~eV~\cite{Shockley}. 
 
Substantial efforts are currently underway to optimise these solar cells by modifying the  
metal-halide network or the organic cation~\cite{Stoumpos2013,Noh2013,Baikie2013,Eperon2014,  
Mosconi2013,Even2013,Even2013b,Umari2013,Eperon2014b}. 
A major challenge in this research area is that, on the experimental front, the structural,  
electronic, and optical properties of metal-organic perovskites are very sensitive to processing conditions.  
Similarly, computational modelling studies appear sensitive to the level of theory and calculation details. 
This state of affairs suggests that it may be convenient to shift the focus from the investigation 
of a few prototypical perovskites to global searches relying on rational design approaches.  
 
In this work we propose that, instead of modifying the metal-halide network as already 
done in Refs.~\cite{Stoumpos2013,Noh2013}, it could be advantageous to start from 
MA-PbI$_3$ (MA = CH$_3$NH$_3$), and modify its electronic and optical properties 
without altering the PbI$_3$ network. 
This strategy would allow us to investigate potential improvements while
retaining highly desirable features such as long carrier lifetimes~\cite{Stranks2013, Xing2013}.
In the following, using a combination  
of geometric arguments and first-principles calculations, we demonstrate that it should
indeed be possible to modulate the optical gap of lead-iodide perovskites over almost
an electronvolt by acting on the Pb-I-Pb bond angles in the PbI$_3$ network. These angles, in turn, can be 
tailored using cations of different sizes, without altering the metal-halide chemistry. 
 
\section*{Results} 
 
\subsection{Platonic model of metal-halide perovskites} 
 
\begin{figure} 
  \begin{center} 
  \includegraphics[width=0.7\textwidth]{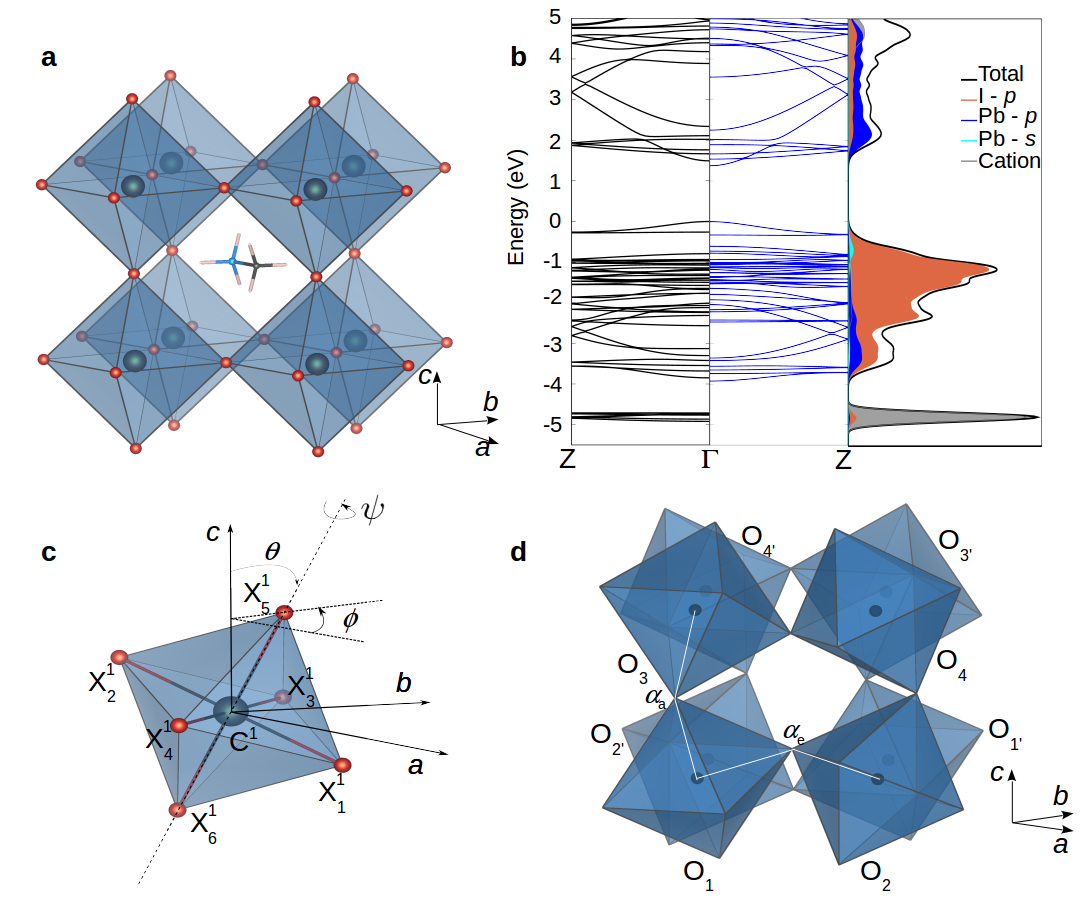} 
  \end{center} 
  \caption{\label{fig1}
  \small 
  \textbf{Platonic model of metal-halide perovskites\hspace{0.2cm}} 
  \textbf{a} Ball-and-stick representation of MA-PbI$_3$ in the high-temperature cubic phase:
  the dark spheres inside the octahedra represent Pb atoms, the red spheres at the octahedral corners
  are I atoms, and the molecule inside the cuboctahedral cavity is CH$_3$NH$_3$.
  \textbf{b} DFT scalar-relativistic electronic band structure and  
  partial density of states of MA-PbI$_3$ in the low-temperature orthorombic phase. 
  In the band structure plot we compare the electronic structure of MA-PbI$_3$ with (left) 
  or without (right) the cation. The density of states shows that the electronic   
  states associated with the cation are located far from the band gap. The top of the valence 
  band is derived from I-$5p$ states and, to a smaller extent, the Pb-$6s$ states. The bottom  
  of the conduction band is derived primarily from Pb-$6p$ and I-$5p$ states.  
  The calculated band gap is direct at $\Gamma$.
  \textbf{c} Schematic representation of the Euler angles defining the rotation 
  of the Platonic octahedron: tilt ($\theta$), 
  spinning ($\psi$), and precession ($\phi$). The convention for labeling the corners 
  of the octahedron is indicated, together with the crystallographic axes.
  \textbf{d} Schematic representation of the orthorhombic unit cell considered in this work 
  consisting of four inequivalent octahedra, O$_1$-O$_4$ (dark blue) and  their 
  periodic replicas O$_1'$-O$_4'$ (light blue). The labelling of the corners
  follows the same convention indicated in (c).
  The intrinsic metal-halide-metal bond angles are highlighted: apical angle
  ($\alpha_{\rm a}$) and equatorial angle ($\alpha_{\rm e}$). The geometric properties of this model
  are derived in the Methods section.}
  \end{figure}

MA-PbI$_3$ belongs to the family of perovskites of chemical formula ABX$_3$. The simplest possible  
ABX$_3$ perovskite consists of a high-symmetry cubic structure belonging to the space group $Pm\bar{3}m$, 
as shown in Fig.~\ref{fig1}a~\cite{Glazer1975}. In this structure the halogen atoms X$=$I occupy the vertices  
of regular corner-sharing BX$_6$ octahedra, while the divalent metal atoms B$=$Pb sit at the centres  
of the octahedra. The smallest volume enclosed by neighboring octahedra defines a cuboctahedral  
cavity and hosts the monovalent cation A$=$CH$_3$NH$_3$. All the other possible perovskite structures  
can be imagined as obtained by rotating or distorting the BX$_6$ octahedra, displacing the 
B metal atoms off-centre, and rotating the A cations within the cuboctahedral cavity~\cite{Glazer1975}. 
 
Fig.~\ref{fig1}b shows the electronic band structure of MA-PbI$_3$, which represents the starting 
point of our investigation. Here we see that the electronic states 
at the top of the valence bands are of halide $p$-character, while those at the bottom of the 
conduction band are mainly derived from the metal $p$ states. At the same time the 
electronic states associated with the cation define bonding and antibonding bands
arranged symmetrically around the band gap and located several electronvolts away from the band edges.  
An analysis of the electronic charge density shows that, as expected, the cation
is singly ionised, and the PbI$_3$ network is negatively charged~\cite{Chang2004,Chiarella2008,Boriello2008}.
These observations suggest that the electronic structure of these metal-organic perovskites may be described 
by removing the cation from the structural model and compensating the negative charge of the PbI$_3$ network 
using a positive background. Fig.~\ref{fig1}b shows that the band structure obtained within this approximation 
is in good agreement with that from a complete calculation. The removal of the cation is especially 
useful since it eliminates the complications associated with the 
observed orientational disorder~\cite{Baikie2013,Stoumpos2013,Kawamura2002,Poglitsch1987}. 
Another important observation can be made by inspecting X-ray diffraction data, as reported  
for example in Ref.~\cite{Baikie2013}. In fact, from the structural parameters reported 
in that work we deduce that the Pb-I bond lengths are extremely regular, the largest relative  
deviation from the average value of 3.18~\AA\ being 0.1\%.  
Furthermore, the diagonals of the PbI$_6$ octahedra are almost perpendicular to each other,  
the maximum deviation from right angles being of 4$^\circ$.  
Therefore, to a good approximation 
the PbI$_6$ octahedra may be considered as Platonic solids carrying the full $O_h$ symmetry. 
 
Based on these considerations we set to build a `Platonic' model of metal-halide perovskites, 
consisting of regular corner-sharing PbI$_6$ octahedra and no cation, as shown in Fig.~\ref{fig1}d. 
In order to construct our model we start from a single regular octahedron in a orthorhombic  
unit cell\cite{Baikie2013}, with one diagonal parallel to the $c$ axis, and the four perpendicular edges  
parallel to the axes $a$ and $b$. In the reference frame of the octahedron this diagonal and its perpendicular 
edges define the apical and equatorial directions, respectively. Within this model, the most general  
perovskite structure is obtained by three successive rotations around $c$, $a$, and $c$ again 
using the standard Euler angles\cite{Goldstein}, as shown in Fig.~\ref{fig1}c: the spinning  
$\psi$ around the apical direction, the tilt $\theta$ around the $a$ axis, and the precession  
$\phi$ around the $c$ axis. After these rotations, the location and orientation of all the remaining  
octahedra in the orthorhombic unit cell are completely determined by the corner-sharing  
connectivity, the metal-halide bond length, and the periodicity of the crystal structure.  
The details of this geometric construction are given in the Methods.  
 
An immediate result of our construction is that the only way to ensure corner-sharing 
connectivity is by having a vanishing spinning angle $\psi$. This leaves us with only two independent angular 
parameters, $\theta$ and $\phi$, which refer to a spherical system of coordinates clamped on 
the crystallographic unit cell. It is convenient to make a change of variables to
intrinsic angular parameters which are independent on the choice of the unit cell. To this end we express  
$\theta$ and $\phi$ in terms of the metal-halide-metal bond angles $\alpha_{\rm a}$ and  
$\alpha_{\rm e}$, as indicated in Fig.~\ref{fig1}d. $\alpha_{\rm a}$ denotes the Pb-I-Pb bond angles  
along the apical direction, and $\alpha_{\rm e}$ those along the equatorial direction.  
The relation between these intrinsic angles and the Euler angles is given in the Methods. 
 
At this point the atomic structure of the Platonic perovskite model is uniquely determined by the choice  
of two metal-halide-metal bond angles and the metal-halide bond length.  
Given that the effect of  bond length variations on the electronic and optical 
properties of these systems has been studied in detail, both theoretically\cite{Boriello2008}  and 
experimentally\cite{Stoumpos2013},
our geometric analysis clearly directs attention to the role of the two inequivalent Pb-I-Pb angles. 
 
\subsection{Band gap and metal-halide-metal bond angle} 
 
\begin{figure} 
  \begin{center} 
  \includegraphics[width=0.85\textwidth]{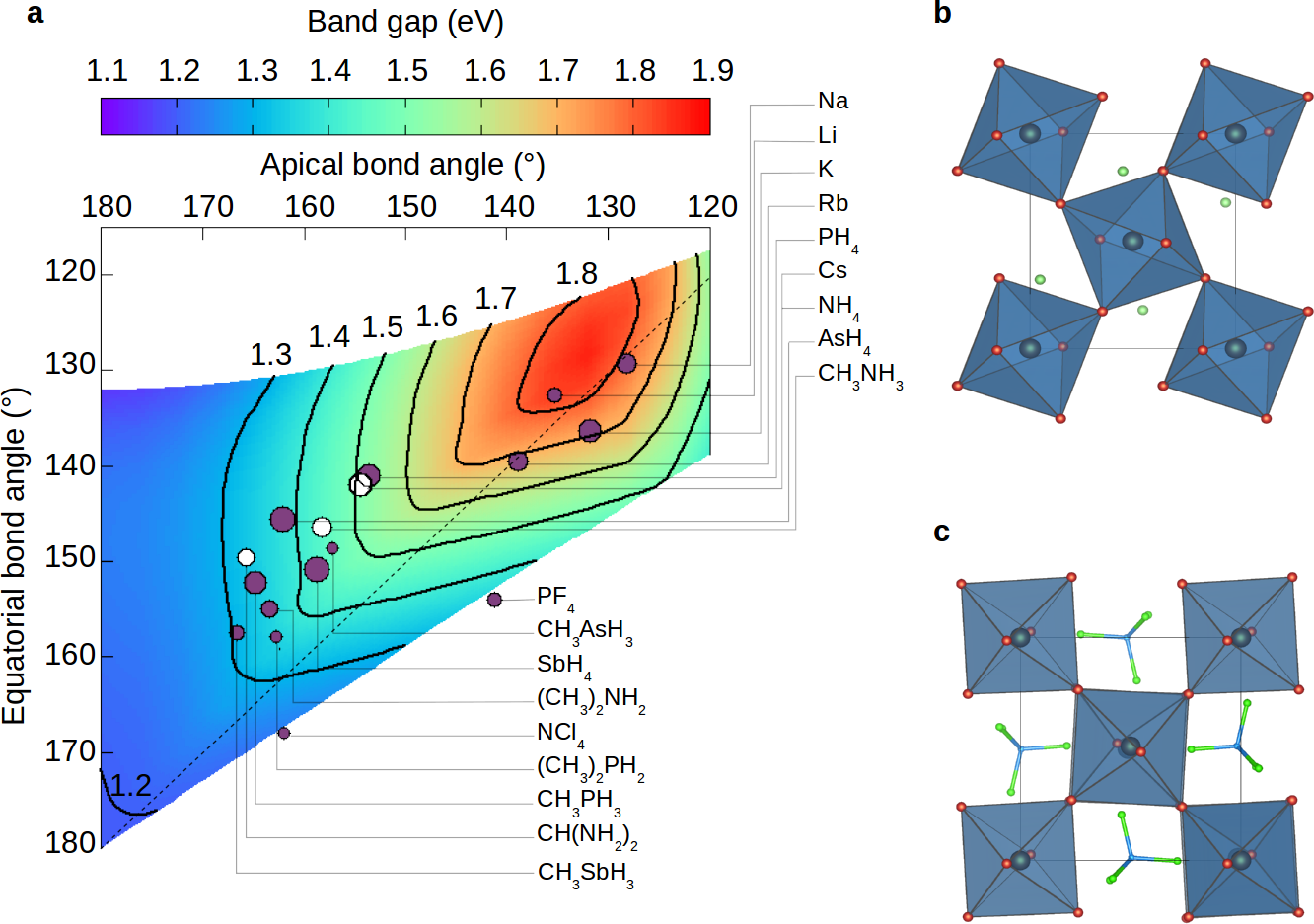} 
  \end{center} 
  \caption{\label{fig2} 
  \small
  \textbf{Mapping the band gap landscape of metal-halide perovskites\hspace{0.2cm}} 
  \textbf{a}  
  Two-dimensional map of the DFT band gap of the Platonic model of PbI$_3$-based perovskites  
  as a function of the apical and equatorial metal-halide-metal bond angles. The Pb-I bond  
  length is set to the experimental value for MA-PbI$_3$~\cite{Baikie2013}. The angles 
  $\alpha_{\rm a}$ and $\alpha_{\rm e}$ are indicated in Fig.~\protect\ref{fig1}d. 
  The calculations for the Platonic model were performed within scalar-relativistic DFT
  (fully-relativistic calculations are shown in Fig.~\protect\ref{fig3}b, see Methods).
  The dashed line at 45$^\circ$ is a guide to the eye and shows the approximate symmetry 
  of the map with respect the the exchange of the apical and equatorial angles. The dark discs represent  
  the angular coordinates $(\alpha_{\rm a},\alpha_{\rm e})$ of realistic models of PbI$_3$-based  
  perovskites including the cation (indicated by the chemical formula), with structures fully optimized   
  within DFT. The apical and equatorial angles are the averages among all the inequivalent 
  angles of the same type. The size of each circle represents the deviation of the Pb-I bond  
  length from that of MA-PbI$_3$ (3.18~\AA). This quantity is displayed because  
  the metal-halide bond length has a small but non-negligible effect on the band gap,  
  which we quantify as a linear shift of $\sim$3~eV/\AA\protect\cite{Boriello2008,Mitzi2005}. 
  For the same angular coordinates, the band gap 
  difference between the largest and the smallest circle in the figure is 0.2~eV. 
  The white discs correspond to perovskites already synthesized~\cite{Baikie2013,Stoumpos2013,Eperon2014}. 
  This map shows that the band gap can be modulated over a much wider range than currently possible. 
  \textbf{b}-\textbf{c} Atomistic models of the hypothetical metal-halide perovskites identified 
  in this work exhibiting the smallest (LiPbI$_3$) and the largest (NCl$_4$PbI$_3$) metal-halide-metal  
  bond angles, respectively. }
  \end{figure}

In Fig.~\ref{fig2} we show the band gap of the Platonic model calculated  
using density functional theory (DFT) as a function of the apical and equatorial bond angles 
$\alpha_{\rm a}$ and $\alpha_{\rm e}$. For definiteness we set the Pb-I bond length to the 
average value measured in Ref.~\cite{Baikie2013}. 
The incomplete coverage of the parameter space in Fig.~\ref{fig2} reflects the fact that 
only certain pairs of apical and equatorial angles are compatible with the corner-sharing 
connectivity, as discussed in the Methods. 
The extremal parameters of this map are, on the one side, $\alpha_{\rm a}=\alpha_{\rm e}=180^\circ$  
(bottom left corner), corresponding to the most symmetric cubic perovskite structure with collinear Pb-I-Pb 
bonds. On the other side we have $\alpha_{\rm a}=\alpha_{\rm e}=120^\circ$ (top right corner), 
which we have chosen by considering a limiting structure where the shortest distance between 
iodine atoms belonging to different octahedra matches the I$_2$ bond length ($\alpha_{\rm a}=0$  
and $\alpha_{\rm e}=126^\circ$). % (2.8~\AA). 

The variation of the angular coordinates in Fig.~\ref{fig2} clearly induces a substantial modulation  
of the band gap, from the top of the mid-infrared (1.1~eV) to the beginning 
of the visible spectrum (1.9~eV). This trend is in line with earlier calculations on simpler two-dimensional  
Sn-I perovskites for solution-processable electronics~\cite{Mitzi2005}. 
 
The qualitative trend shown in Fig.~\ref{fig2} can be interpreted using elementary tight-binding  
arguments. If we consider for definiteness the bottom of the conduction band at $\Gamma$, which is 
most affected by bond angles, in the case of $\alpha_{\rm a}=\alpha_{\rm e}=180^\circ$ there are  
three degenerate electronic states derived from the $6p$ orbitals of Pb (without considering
spin-orbit coupling for simplicity; this coupling is fully included in all our final results 
presented in Fig.~\ref{fig3}b). It is intuitive that the energy of these states  
results mainly from $pp\sigma$ bonding integrals since all the Pb-I bonds are collinear~\cite{Slater}. 
By moving away from the bottom left corner, the degeneracy is lifted  
as the bond integrals acquire $pp\pi$ components weighted by the angles $\alpha_{\rm a}$  
and~$\alpha_{\rm e}$. Both components tend to raise the energy levels, therefore the conduction  
band bottom results from the competition between the antibonding orbitals with the lowest energy 
along the apical or the equatorial direction. This is consistent with the approximate symmetry 
of the map in Fig.~\ref{fig2} with respect to the line corresponding to 
$\alpha_{\rm a}=\alpha_{\rm e}$. More importantly this reasoning suggests that the band gap of the Platonic 
perovskite model is governed by the largest Pb-I-Pb bond angle in the metal-halide network,
consistent with the trend in Fig.~\ref{fig2}.
This relation between band gap and bond angles
holds unchanged when considering fully relativistic calculations including spin-orbit corrections.
In fact, as shown in Supplementary Figure~1, relativistic effects induce a large but slowly-varying
red shift of 0.8-1.1 eV across the family of compounds considered in this work.
For the sake of completeness we also checked that excitonic effects play only a minor role, with
exciton binding energies never exceeding 80~meV as shown by Supplementary Figure~2. 
Details on the calculation of
exciton binding energies are given in the Supplementary Methods, and the results are
discussed in the Supplementary Note~1.
 
After taking into account the presence of the cation and the variations in the Pb-I bond lengths  
and Pb-I-Pb angles, the calculated band gaps  
deviate from the idealised predictions in Fig.~\ref{fig2} by 0.1-0.3~eV (see Fig.~\ref{fig3}  
and discussion below). Nonetheless the general trend, which spans a range of up to~1~eV,  
is robust. This picture is also confirmed by relativistic $GW$ calculations on a few selected 
compounds, as shown in Supplementary Figure~4.
The trend identified here suggests that in order to make low-gap perovskites for  
optimum photovoltaic efficiency we need to engineer structures with minimal octahedral tilt. 
On the other hand, in order to make large-gap perovskites for light-emitting diodes operating in 
the visible, we need to design structures with maximum tilt. The remaining question is 
how the tilt angle can be controlled. 

\subsection{Controlling metal-halide-metal bond angles via the steric size of the cation}  
 
The Pb-I-Pb bond angles define the volume of the cuboctahedral cavity, therefore it is intuitive 
that the use of larger cations may lead to values of $\alpha_{\rm a}$ and $\alpha_{\rm e}$ closer  
to straight angles\cite{Mitzi-book}. In the search for such cations we perform DFT calculations  
for existing structures as well as for many hypothetical structures not considered hitherto, including full  
structural optimisations (atomic positions, unit cell, and lattice parameters) and spin-orbit interactions, 
in the presence of the cation (see Methods).  
In order to relate these structures and their properties to the predictions of the Platonic model, we  
consider the average values of the Pb-I-Pb bond angles and of the Pb-I bond length over a crystalline unit  
cell. These values are used as the coordinates needed for locating each structure on the map of Fig.~\ref{fig2}.  
 
Previously reported cations are methylammonium CH$_3$NH$_3^+$ \cite{Baikie2013},  
formamidinium CH$_2$(NH$_2$)$_2^+$~\cite{Eperon2014, Stoumpos2013}, the alkali metal Cs$^+$~\cite{Eperon2014} 
(synthesised) and ammonium NH$_4^+$~\cite{Walsh2013,Walsh2014, Walsh2014b} (proposed).  
As shown in Fig.~\ref{fig2} all these cations cluster  
around the centre of the map, therefore they are not expected to yield a significant modulation  
of the band gap. In line with this result their measured optical absorption onsets are very similar,  
within 0.25~eV~\cite{Eperon2014,Stoumpos2013}. 
 
In order to explore a wider portion of the map we consider four different families of cations 
generated from those above. The first family consists of secondary, tertiary, and quaternary ammonium  
cations, namely di-, tri- and tetra-methylammonium. These are large molecules 
obtained by replacing hydrogen atoms bonded to the N atom with methyl groups. 
The second family is generated from ammonium by descending the pnictogen column in the periodic table, 
that is by substituting N for P, As, or Sb. Members of this family include phosphonium (PH$_4^+$), arsonium 
(AsH$_4^+$), stibonium (SbH$_4^+$), and similarly the methylammonium analogues  
CH$_3$PH$_3^+$, CH$_3$AsH$_3^+$ and CH$_3$SbH$_3^+$. The third family of cations that we consider 
is obtained by replacing hydrogen in ammonium by halogen atoms, and possibly nitrogen by another 
pnictogen: in this group we have NF$_4^+$, NCl$_4^+$, PF$_4^+$, and PCl$_4^+$. 
The fourth and final family simply consists of the alkali metals Li$^+$, Na$^+$, K$^+$, and Rb$^+$. 
In total we consider 22 structures, including 18 hypothetical 
compounds not reported to date. 
 
In order to quantify the steric size of each cation, we use the radius of the sphere which 
contains 95\% of the DFT electron density. This choice ensures that the steric radii of the 
alkali metals are in agreement with their standard ionic radii \cite{Shannon}. For completeness the calculated  
steric sizes are reported in Supplementary Table~1. Fig.~\ref{fig3} shows that the steric radii
of the cations considered in this work span a wide range, from 0.7~\AA\ (Li$^+$) to 2.9~\AA\ (PCl$_4^+$). 
 
Our structural optimisations indicate that for some of the large cations the three-dimensional  
perovskite network is significantly distorted. This is in line with previous studies showing that 
large molecular cations determine a reorganisation of the three-dimensional (3D) perovskite network into 
a two-dimensional (2D) layered structure.\cite{Mitzi1995, Tanaka2005} These large molecules include tertiary  
and quaternary methylammonium cations, as well as the tetrafluoride NF$_4^+$ and the tetrachloride PCl$_4^+$.  
As these structures depart substantially from the three-dimensional metal-halide network considered in this 
work, a separate analysis would be required. Accordingly we do not consider them further.  
 
Fig.~\ref{fig3}a shows that the largest Pb-I-Pb angles correlate strongly with the steric size of 
the cation (Spearman correlation 88\%), in line with our initial expectation. 
The newly proposed structures span a range of angles from 130$^\circ$ to 170$^\circ$, thereby covering 
a much wider portion of the band gap map in Fig.~\ref{fig2} than presently possible. 
In addition Fig.~\ref{fig3}b shows that, in agreement with the predictions of the Platonic model,  
the band gap correlates strongly with the largest Pb-I-Pb angle in the unit cell
(Spearman correlation 91\%). 
For completeness in Fig.~\ref{fig3}b we report the band gaps obtained from this model as well as 
those calculated using the fully optimised structures within fully-relativistic DFT calculations. 
The scatter of the data which is clearly visible in Fig.~\ref{fig3}b is mostly due to the
variation of the Pb-I bond lengths in the metal-halide network. 
We calculated the correlation coefficient between band gap and bond lengths using the same data,
and we obtained a very weak correlation (Spearman coefficient 9\%). This test clearly
indicates that the role of bond lengths represents a second-order effect, thereby providing
further support to our Platonic model.
Finally,  
Fig.~\ref{fig3}c shows our main finding, namely that by suitably choosing the cation 
it is possible, at least in principle, to fine-tune the band gap almost continuously over a 
very wide range of photon energies.
 
Our analysis highlights several new promising structures, in particular 
perovskites with phosphonium (PH$_4^+$), arsonium (AsH$_4^+$), and stibonium (SbH$_4^+$),
all with band gaps ranging between 1.2 and 1.4~eV. These values fall in the middle of  
the range for optimum photovoltaic  
efficiency, and suggest that our new structures hold potential as novel solar cells materials. 
Besides the low gap, these perovskites are expected to exhibit higher mobilities than
MA-PbI$_3$. In fact, as shown in Supplementary Figure~2a and 2b, the carrier
effective masses decrease towards larger bond angles, consistent with the concomitant reduction
of the band gap. In addition to lower gaps and effective masses, 
large cations have the advantage that the filling of
the cuboctahedral cavity should counter the known tendency of perovskite solar cells to degrade 
by water incorporation~\cite{Noh2013}.
 
In order to evaluate the practical feasibility of the hypothetical perovskites identified
in this work we searched for possible dynamical instabilities in one of the structures
with the smallest band gap, AsH$_4$PbI$_3$ (NCl$_4^+$ is less interesting due to safety 
concerns related to chlorides). This structure is especially interesting since
the preparation could proceed through the corresponding iodide AsH$_4$I~\cite{ash4i},
similarly to the case of methylammonium iodide CH$_3$NH$_3$I.
As shown by Supplementary Figure~3, the vibrational density of states calculated
using density-functional perturbation theory~\cite{Baroni} does not exhibit any soft modes,
thereby indicating that the 3D perovskite structure of AsH$_4$PbI$_3$ 
should be stable against distortions.
More generally, in order to assess the stability of all the structures considered here,
we move to a simple semi-empirical approach based on the Goldschmidt tolerance factor $t$~\cite{Goldschmidt,
Mitzi-book} 
(see Methods for the definition of $t$). It is known empirically that 3D perovskites should 
form when the tolerance factor falls within a narrow range $t=0.7-1.1$~\cite{Goldschmidt,
Mitzi-book,Jones-book,Green2014}. 
Fig.~\ref{fig4}a shows that the tolerance factors derived from our calculated steric sizes
falls within this range for most of the structures considered here.
In particular, according to this criterion, all the cations up to Rb and K should be stable.

In order to demonstrate the band gap tunability we attempted the synthesis of perovskites
based on elemental cations, building on our experience with CsPbI$_3$~\cite{Eperon2014} (see Methods).
As deposited, we observed that CsPbI$_3$ formed a pale phase with a wide 
bandgap. Upon annealing this structure turned into a dark phase with a band gap of 1.73 eV, 
which we identify as the 3D perovskite structure discussed in this work. The synthesis of films 
with cations smaller than Cs (Rb, K, Na, Li) yielded similar pale phases. 
However, upon annealing, some of the films degraded via sublimation before we could observe 
a dark phase as for CsPbI$_3$, suggesting that 3D perovskites were not formed. This finding 
is in agreement with our analysis of the Goldschmidt tolerance factor shown in Fig.~\ref{fig4}a.
While the proposed 3D perovskites could possibly be synthesised using
higher pressures or longer annealing runs at lower temperatures, here for the sake of simplicity
we did not pursue this direction. Instead we considered an 
alternative approach and explored mixtures of cations, as suggested in Ref.~\cite{Trots2008}: if mixtures of Cs and smaller 
cations could gradually shift the bandgap towards the larger values, this would confirm 
our predicted tunabiliy via steric effects.

As a proof-of-concept we prepared perovskites with Rb/Cs cation mixtures, namely
Rb$_x$Cs$_{1-x}$PbI$_3$ with $x=0$, $x=0.1$, and $x=0.2$ (never hitherto reported). Optical characterisation
of these films, shown in Fig.~\ref{fig4}b, indicates that the absorption edge blue-shifts from
720~nm to 690~nm in going from Cs to Rb. This result is in excellent agreement
with our calculations, which predict a corresponding blue shift of 40~nm.

Our findings are also in line with a recent study (which appeared while the present manuscript
was under review) where the authors demonstrated band gap tunability via mixing of
CH$_3$NH$_3^+$/Cs$^+$ mixtures~\cite{Choi2014}.
Taken together our present results and those of Ref.\cite{Pellet2014,Eperon2014, Stoumpos2013} confirm the tunability
of the band gap via the steric size of the cation over almost half of our predicted range,
thereby providing strong support to our theory.

\section*{Discussion} 
 
In conclusion, by means of first-principles computational design we have established that 
it should be possible to continuously tune the band gap  
of lead-iodide perovskites over almost one electronvolt by controlling the octahedral 
tilt via the steric size of the cation. We identified the largest metal-halide-metal bond angle 
as a reliable descriptor of the band gap in these systems. Following the predictions of a 
geometric model we investigated many new potential candidate perovskites retaining 
the metal-halide network of methylammonium lead-iodide, including fourteen structures not
reported to date. 
Prompted by our theoretical analysis, we proceeded to the experimental synthesis and 
characterization of novel mixed-cation
Rb$_x$Cs$_{1-x}$PbI$_3$ perovskites, and we confirmed our predicted band gap trends
via optical spectroscopy.

Since our theory is based on very general considerations and its predictions are not linked 
to the underlying calculation method, it is expected that it will carry similar predictive power 
across other families of metal-halide perovskites. More fundamentally, the ability to control  
the band gap via the steric size of the cation, as proposed in this work, may become a  
new paradigm in the solution-deposition of solar cells, light-emitting diodes, and photonic  
structures with layer-by-layer control of their optical properties.  

\newpage 
\section*{Methods} 
\section*{\small Computational methods} 
 
All calculations were performed within the local density approximation to density 
functional theory~\cite{Ceperley1980,Perdew1981} using planewaves and  
pseudopotentials, as implemented in the {\tt Quantum ESPRESSO} software package~\cite{QE-2009}.  
Only valence electrons were described explicitly, and the core-valence interaction 
was taken into account by means of ultrasoft pseudopotentials~\cite{Rappe1990,Vanderbilt}
with non-linear core correction~\cite{Louie}. 
For Pb and I we used fully relativistic  
ultrasoft pseudopotentials. In the case of Pb we included the $5d$ semicore states,  
and for Sb, Rb, and Cs we used norm conserving pseudopotentials~\cite{Troullier}. 
The electron wavefunctions 
and the charge density were expanded in planewaves basis sets with kinetic energy cutoffs 
40~Ry and 200~Ry, respectively. Structural optimization was performed by sampling the Brillouin 
zone using uniform and unshifted 4$\times$4$\times$4 meshes. 
Electronic structure calculations were performed using a 6$\times$6$\times$6  
Brillouin zone grid. All structural optimizations were initialized using the crystallographic 
data for MA-PbI$_3$ reported in Ref.~\cite{Baikie2013}. The initial coordinates of the molecular 
cations were chosen in such a way that the centre of mass is located in the middle of the cuboctahedral  
cavity, and the orientation of the four inequivalent cations in one unit cell follows the 
symmetry properties of the $Pnma$ space group. 
Given the complexity of the total energy landscape of these systems, we checked the
sensitivity of the structure with respect to the choice of the initial configuration
in the case of NH$_4$PbI$_3$. This compound lies close to the stability limit (Figure~4b),
and can form quasi-1D structures in the presence of water~\cite{Bedlivy, Fan}.  
In this case we performed structural optimizations using three additional initial configurations:
(i) $\alpha_{\rm a} = \alpha_{\rm e} = 0$, (ii) $\alpha_{\rm a} = 0$ and $\alpha_{\rm e}\neq 0$
with octhaedra in phase along the apical direction, (iii) same as in (ii) but with octahedra
out of phase. The optimized structures are all 3D perovskites, with total energy differences within 6~meV/atom.
The largest deviation from the band gap reported in Fig.~3c is of 0.2~eV, and reflects the
differences in the bond angles, following the trend predicted by the Platonic model in Fig.~2.

In the case of MA-PbI$_3$ it has been established that the spin-orbit 
interaction tends to reduce the scalar-relativistic band gap by approximately 
1-1.1~eV~\cite{Umari2013,Even2013}. This is consistent with our results shown in Fig.~\ref{fig3}b. 
On the other hand, $GW$ quasiparticle corrections do increase the value of the band gap~\cite{Umari2013}.
Given these compensating corrections,
and the fact that the very small relativistic band gaps are likely to induce significant 
errors in the calculation of the screened Coulomb interaction $W$, 
it is sensible to take scalar-relativistic calculations as the most consistent and unambiguous
way to capture trends and identify the most promising structures. In order to check that the band gap trend
obtained with DFT/LDA holds when using a higher level theory, we performed $G_0W_0$ calculations using the
\texttt{Yambo} code~\cite{Marini2009} for a few selected structures, namely AsH$_4$PbI$_3$, CH$_3$NH$_3$PbI$_3$, 
CsPbI$_3$ and LiPbI$_3$. Full details of these calculations are given in the Supplementary Methods and 
Supplementary Note 2.

\section*{\small Platonic model of metal-halide perovskites}

Our ideal model of ABX$_3$ metal-organic halide perovskites consists of four inequivalent BX$_6$
Platonic octahedra in an orthorombic unit cell~\cite{Baikie2013}, in absence of the cation.
The octahedra are labelled O$_1$-O$_4$, as indicated in Fig.~\ref{fig1}d, and are connected
in a corner-sharing configuration. The B atoms occupy the centres of the octahedra, the X atoms
are located at the corners, and the metal-halide B-X distance is denoted by $d$.
We set the origin of the reference frame at the centre of the octahedron O$_1$, with its axes
along the primitive vectors of the crystal lattice. The centres of the remaining three octahedra
O$_2$-O$_4$ are denoted by $(x, y, 0)$, $(0, 0, z)$, and $(x, y, z)$, respectively.
The coordinates $x$, $y$, and $z$
will be determined below starting from the Euler angles described in the main text.
The lattice parameters are expressed in terms of the centres of the octahedra as
$a = 2x$, $b = 2y$ and $c=2z$. The corners of the octhedra are labelled as X$_i^j$, with
the superscript $j$ identifying the octahedron, and the subscript $i$ identifying a corner of
the octahedron, as shown in Fig.~\ref{fig1}c. The coordinates of each corner X$_i^j$
are referred to as $(x_i^j, y_i^j, z_i^j)$.

{\it Octahedron {\rm O}$_\mathbf{1}$.}
We consider an initial configuration where the octahedron O$_1$ has its apical diagonal along
the $c$ axis, and its equatorial edges along the $a$ and $b$ axes. The coordinates of the corners
X$_1^1$, X$_3^1$, and X$_5^1$ are by $(1, 1, 0)d/\sqrt{2}$, $(-1,1,0)d/\sqrt{2}$, and $(0,0,1)d$,
respectively, while  X$_2^1$, X$_4^1$, and X$_6^1$ are obtained from these by changing the
sign of all coordinates.
In order to generate the most general structural models we perform a rotation of O$_1$
according to the three Euler angles $\psi$, $\theta$, and $\phi$. These angles define
a sequence of spin, tilt, and precession around the $c$ axis. The tilt $\theta$ is performed
via a rotation around $a$. This is illustrated in Fig.~1c.
Following Ref.~\cite{Goldstein} the matrix defining the rotation of the octahedron is:

 $$
 \left(
 \begin{array}{ccc}
 \cos{\psi}\cos{\phi}-\cos{\theta}\sin{\phi}\sin{\psi} & 
 -\sin{\psi}\cos{\phi}-\cos{\theta}\sin{\phi}\cos{\psi}&
 \phantom{-}\sin{\theta}\sin{\phi} \\
 \cos{\psi}\sin{\phi}+\cos{\theta}\cos{\phi}\sin{\psi} & 
 -\sin{\psi}\sin{\phi}+\cos{\theta}\cos{\phi}\cos{\psi}&
 -\sin{\theta}\cos{\phi}\\
 \sin{\theta}\sin{\psi} & 
 \sin{\theta}\cos{\psi} & 
 \cos{\theta}
 \end{array} 
 \right).
 $$

By applying this rotation to the initial coordinates of O$_1$ we obtain the position of the corners
X$_i^1$ ($i=1,\dots,6$) in their final orientation. These new coordinates are used to locate
the centres and orientation of the remaining three octahedra.

{\it Octahedron {\rm O}$_\mathbf{2}$}
In order to determine the centre $(x,y,0)$ of the octahedron O$_2$ we use the following
observations: O$_1$ shares the corner X$_1^1$ = X$_2^2$ with O$_2$, and the corner X$_3^1$ = X$_4^{2'}$
with O$_{2'}$, the periodic replica of O$_2$ along the $a$ axis. Furthermore
O$_2$ shares the corner X$_3^2$ = X$_4^{1'}$ with O$_{1'}$, the periodic replica of O$_1$ along
the $b$ axis.
By requesting that the distance between the centre (C$^i$) of each octahedron and each corner is $d$
in all cases, and that O$_2$ and O$_{2'}$ are both regular, we obtain the following equations
relating the centre of O$_2$ with the position of the corner X$_1^1$:
  $$
  \begin{array}{rl}
  \mbox{C}^1 \mbox{X}_1^1= \mbox{C}^2 \mbox{X}_1^1            :~& x^2+y^2-2xx_1^1-2yy_1^1 = 0, \\
  \mbox{C}^{1'}\mbox{X}_4^{1'}=\mbox{C}^2 \mbox{X}_4^{1'}     :~& x^2+y^2+2xx_3^1-2yy_3^1 = 0, \\
  \mbox{C}^{2}\mbox{X}_4^{1'}\perp \mbox{C}^2 \mbox{X}_1^{1}  :~& x^2-y^2+x(x_3^1-x_1^1)+y(y_1^1+y_3^1) = 0.
  \end{array}
  $$
After expressing the coordinates of X$_1^1$ in terms of the Euler angles $\psi$, $\theta$, and $\phi$,
we find that the above relations are verified simultaneously only when the spinning angle vanishes,
$\psi=0$. In this case the coordinates of the centre of O$_2$ are given by
$ x = d\sqrt{2} \cos{\phi}$, $y = d\sqrt{2} \cos{\theta}\cos{\phi}$,

and the coordinates of its corners are determined using the condition that the octahedron be regular.
The explicit expressions for the coordinates of O$_2$ in terms of the corners of O$_1$ are:

\begin{tabular}{llllllll}
X$_1^2$: & $(2x-x_1^1, 2y-y_1^1, -z_1^1)$ && X$_2^2$: & $(x_1^1, y_1^1, z_1^1)$ &&
     X$_3^2$: & $(x_4^1, y_4^1+2y, z_4^1)$ \\
X$_4^2$: & $(2x-x_4^1, -y_4^1, -z_4^1)$ &&  X$_5^2$: & $(x+p,  y+q,  r)$ &&
     X$_6^2$: & $(x-p,  y-q, -r)$
\end{tabular}

where the auxiliary variables $p$, $q$, $r$ are given by:

$p = \displaystyle\frac{(y_4^1+y)z_1^1-z_4^1(y_1^1-y)}{d}, \,
q = \displaystyle\frac{(x_1^1-x)z_4^1-z_1^1(x_4^1-x)}{d}$, \\
$r = \displaystyle\frac{(x_4^1-x)(y_1^1-y)-(y_4^1+y)(x_1^1-x)}{d}$.

From the location of O$_2$ can can also determine the lattice parameters:
$a = 2d\sqrt{2}\cos{\phi}$, $b = 2d\sqrt{2}\cos{\theta}\cos{\phi}$, and $c = 4d\cos{\theta}$,
where the last equation follows simply from the common tilt angle of O$_1$ and O$_2$
and the fact that they represent half of the unit cell along the $c$ axis.

In the case of vanishing tilt angle, $\theta=0$, the spinning angle
and the precession angle identify the same rotation, hence their distinction is not meaningful.
In this case the relevant parameter appearing in all the equations is their sum $\phi+\psi$.
For consistency with the above rule that $\psi=0$ whenever $\theta\ne 0$, we here adopt the
convention that $\psi=0$ always. In this case, when $\theta=0$ the only other independent
parameter is the precession angle $\phi$. Obviously this choice does not affect our results
in any way.

{\it Octahedra {\rm O}$_\mathbf{3}$ and {\rm O}$_\mathbf{4}$.}
From Fig.~\ref{fig1}d we see that octahedron O$_3$ shares the corner X$_5^1$ = X$_6^3$ with
O$_1$. As a consequence we have that octahedra O$_3$ and O$_4$ are centered at the positions
$2d(0,0,\cos{\theta})$ and $d\sqrt{2}(\cos{\phi},\cos{\theta}\cos{\phi}, \sqrt{2}\cos{\theta})$,
respectively. The corners of these octahedra are readily obtained by taking into account
their regularity and the corner-sharing connectivity between O$_3$ and O$_4$ and between
O$_2$ and O$_4$.

From the relations stated here it is clear that a cubic perovskite is obtained
in the special case $\theta=\phi=0$. Furthermore, when the tilt angle vanishes, $\theta=0$,
we obtain two inequivalent tetragonal structures with identical lattice parameters.
These structures correspond to the situations where the spinning angles of
neighboring octahedra along the apical direction are in-phase (same magnitude and same sign)
or out-of-phase (same magnitude but different sign).

As a general remark, the fact that the structure of the Platonic model is completely defined once
we specify the values of two angles and the B-X bond length is quite remarkable. This property
simplifies considerably the task of rationalizing the physics of metal-organic halide perovskites.
Furthermore the model illustrated here carries general validity and it can easily be adapted
to describe other idealized perovskite structures.

{\it Apical and equatorial bond angles.}
The Euler angles $\theta$ and $\phi$ can be related to the intrinsic B-X-B bond angles.
To this end we observe in Fig.~\ref{fig1}d that $\alpha_{\rm e}$ is the angle
between O$_1$ and O$_2$ through the corner X$_1^1$, and $\alpha_{\rm a}$ is the angle
between O$_1$ and O$_3$ through X$_5^1$. Using the coordinates of the centres of O$_1$-O$_3$
and the corners X$_1^1$ and X$_5^1$ determined as above, we find immediately
$\cos\alpha_{\rm a} =1-2\cos^2\theta$ and $\cos\alpha_{\rm e}=1-\cos^2\phi\,(1+\cos^2\theta)$.
Since $\cos^2(\phi)\le 1$ the last two relations can be combined to yield the constraint 
$\cos (\pi-\alpha_{\rm a}) \ge 2\cos (\pi-\alpha_{\rm e}) -1$. For all the angles considered  
in Fig.~\ref{fig2} the expansion of this inequality to first order in the angles is accurate  
to within 3\% and reads $\pi-\alpha_{\rm e} \ge (\pi-\alpha_{\rm a})/\sqrt{2}$. As a result, 
the Platonic model does not admit solutions with continuous corner-sharing connectivity  
in a region of the map bound by a straight line, as can be seen in Fig.~\ref{fig2}.

\section*{\small Tolerance Factor}

The Goldschmidt tolerance factor is given by $t = (R_{\rm A}+R_{\rm X})/\sqrt{2}
(R_{\rm B}+R_{\rm X})$, with $R_{\rm A}$, $R_{\rm B}$, and $R_{\rm X}$ the ionic radii of the
elements in ABX$_3$ perovskites~\cite{Goldschmidt}.
It was found empirically that 3D perovskite
form when $t$ is in the range 0.7-1.1~\cite{Goldschmidt, Mitzi-book, Jones-book, Green2014}. 
The lower bound corresponds to a situation where the cation is close enough to 
the halogen component to form a bond, thereby preventing the formation of a perovskite 
structure. The upper bound describes the case of the close-packed cubic perovskite structure.
For values of $t$ larger than 1 the perovskite structure is expected to distort 
or even form a layered perovskite structure~\cite{Im2012}.
For each structure we evaluated the Goldschmidt tolerance factor by using the steric sizes calculated 
via the DFT electron density and reported in Supplementary Table~1. 

\section*{\small Materials synthesis and characterization}

{\it Perovskite precursor preparation}: APbI$_3$ precursors were prepared by dissolving equimolar 
amounts of AI (where A = Li, Na, K, Rb, Cs) and PbI$_2$ in N,N-dimethylformamide at 0.5M, 
in a nitrogen-filled glovebox. Mixtures were made by simply mixing the prepared precursors 
in the desired ratio.\\
{\it Film formation}: Glass substrates were cleaned with acetone, isopropanol, and oxygen plasma 
treatment. Thin films were formed by spin-coating on the substrates at 2000~rpm in the glovebox, 
and subsequently annealed on a hotplate by increasing the temperature until the formation of the dark phase 
was observed. This occurred at $\sim$400$^\circ$C for CsPbI$_3$ and even higher for the 
Cs/Rb mixtures discussed in the main text. The 11\% Rb film changed colour at ~420$^\circ$C and 
the 20\% Rb film at ~470$^\circ$C.\\
{\it Optical measurements}: Absorbance spectra were collected with a Varian Cary 300 UV-Vis 
spectrophotometer with an internally coupled integrating sphere.

\section*{Note added during revision}
While this work was under peer review a related work exploring the relation between the size of
cations and the electronic and optical properties of metal-halide perovskites was submitted 
and published~\cite{Amat2014}. The key novelty of the present work 
is that we develop a predictive universal theory of band gap tuning via steric effects,
and we perform experiments which confirm {\it a posteriori} our predictions.

\newpage 
 
\begin{addendum} 
\item[Acknowledgements] 
The authors wish to thank A. N. Kolmogorov for fruitful discussions.
This work was supported by the European Research Council (EU FP7 / ERC grant no. 239578),  
the UK Engineering and Physical Sciences Research Council (Grant No. EP/J009857/1) and  
the Leverhulme Trust (Grant RL-2012-001). 
G.E.E. is supported by the UK EPSRC
and Oxford Photovoltaics Ltd. through a Nanotechnology KTN CASE award.
Calculations were performed at the Oxford Supercomputing  
Centre and at the Oxford Materials Modelling Laboratory. All structural models were rendered using
VESTA\cite{VESTA}.
 
\item[Author contributions] 
M.R.F. performed the computational research and analysed the results. 
G.E.E. and H.J.S. performed the experimental synthesis and characterisation.
F.G. designed the research and led the project. All 
authors participated in the preparation of the manuscript. 
 
\item[Additional information]  
Supplementary information accompanies the paper online. Correspondence and requests for materials 
should be addressed to F.G. (feliciano.giustino@materials.ox.ac.uk). 
 
\item[Competing financial interests] The authors declare that they have no competing financial interests. 
\end{addendum} 
 
  \begin{figure} 
  \begin{center} 
  \includegraphics[width=0.9\textwidth]{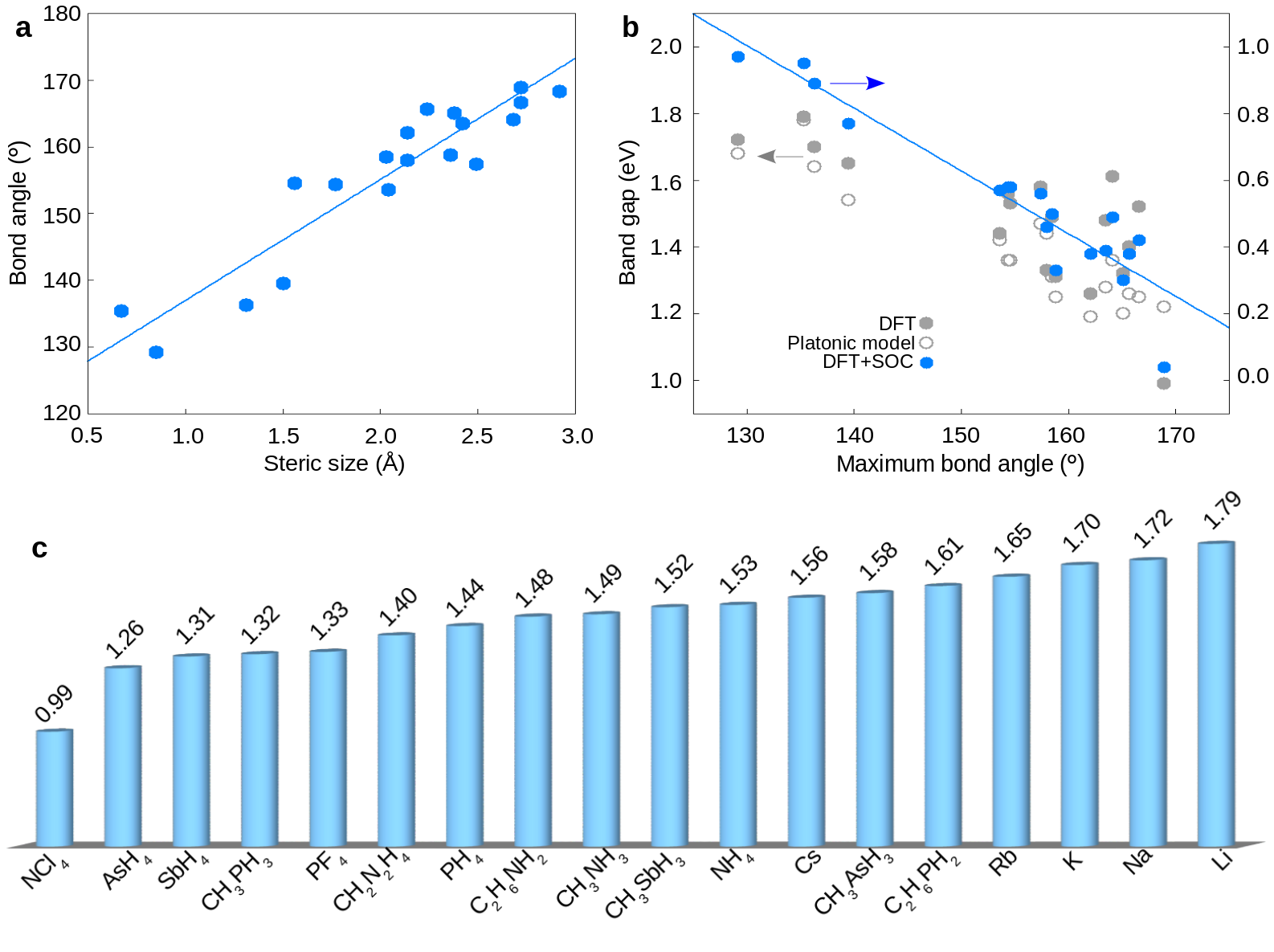} 
  \end{center} 
  \caption{\label{fig3} 
  \small
  \textbf{Tuning the band gap of metal-halide perovskites via the steric size of  
  the cation\hspace{0.2cm}} 
  \textbf{a}~Correlation between the apical and equatorial bond angles of PbI$_3$-based  
  perovskites and the steric radius of the cation (discs). As the size of the cation 
  increases the metal-halide-metal bonds (see Fig.~\protect\ref{fig1}d) tend to become  
  collinear. A linear least-square fit to the data (straight line) yields a slope of 18$^\circ$/\AA.  
  \textbf{b} Correlation between the DFT band gap and the largest metal-halide-metal bond 
  angle in the structure. We show calculations for the Platonic model (circles) and  
  fully optimized structures within scalar-relativistic (grey discs) and fully-relativistic (blue discs)  
  DFT. The arrows point to the scales corresponding to each set of data. 
  The significant data dispersion at large angles is due to the distortions of the octahedra.
  The calculations based on the Platonic model include the effect of the average
  Pb-I bond length on the band gap, as discussed in Fig.~\protect\ref{fig2}.
  \textbf{c} Calculated band gaps (in eV) of all the PbI$_3$-based perovskites considered in this 
  work. The band gaps were obtained after full structural optimization within scalar relativistic DFT 
  (see Methods). 
  } 
  \end{figure} 
 
  \begin{figure} 
  \begin{center} 
  \includegraphics[width=\textwidth]{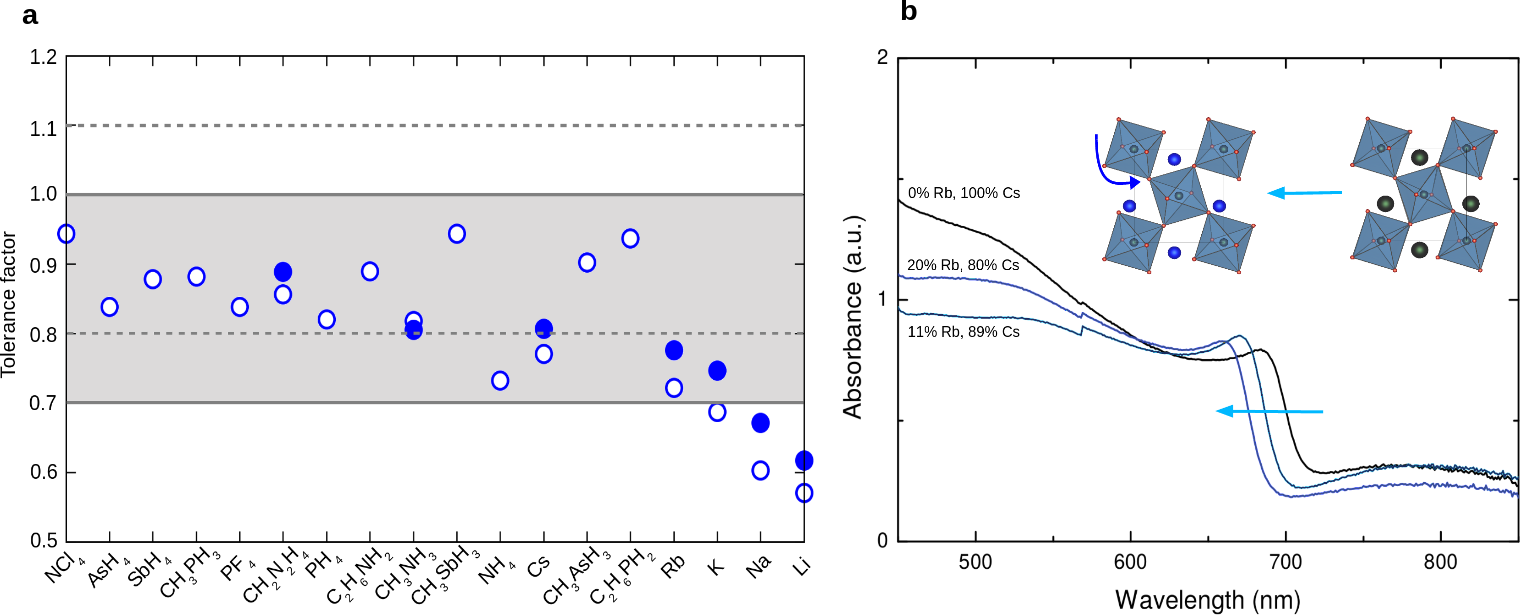} 
  \end{center} 
  \caption{\label{fig4} 
  \textbf{Materials stability and optical measurements\hspace{0.2cm}} 
  \textbf{a}~Calculated Goldschmidt tolerance factors for all the perovskite structures considered in this
  work. The open circles are obtained using the steric sizes calculated from DFT (Supplementary Table~1). 
  For comparison we also show the tolerance factors obtained by using the ionic radii reported in Ref.~\cite{Shannon}
  (filled circles). The grey area bound by the solid grey lines correspond to the geometric criterion
  for the Goldschmidt tolerance factor~\cite{Jones-book, Goldschmidt}, while the dashed lines correspond to the 
  empirical range proposed in Ref.~\cite{Green2014}.
  \textbf{b}~Measured absorbance spectra of the dark phase of Rb$_x$Cs$_{1-x}$PbI$_3$ 
  perovskite thin films, with $x=0$, $x=0.1$, and $x=0.2$. When increasing the
  Rb content a continuous blue shift of the absorption onset is observed,
  from 720~nm ($x=0$) to 690~nm ($x=0.2$). This blue shift corresponds to an increase in band gap, 
  and is assigned to the more pronounced octahedral tilt in the PbI$_3$ network,
  as illustrated in the inset.
  }
  \end{figure} 

\newpage

\title{{\bf Steric engineering of metal-halide perovskites\\ with tunable optical band gaps\vspace{0.3cm}\\
Supplementary Information}}

\author{{\bf Marina R. Filip$^1$, Giles E. Eperon$^2$, Henry J. Snaith$^2$ \& Feliciano Giustino$^1$}} 
 
\maketitle 

\vspace{0.2cm} 
\begin{affiliations} 
\item[$^1$] Department of Materials, University of Oxford, Parks Road, Oxford, OX1 3PH 
\vspace{-0.2cm} 
\item[$^2$] Department of Physics, University of Oxford, Clarendon Laboratory, Parks Road, Oxford, OX1 3PU, UK
\end{affiliations}

  \begin{figure}[p]
  \begin{center}
  \includegraphics[width=0.6\textwidth]{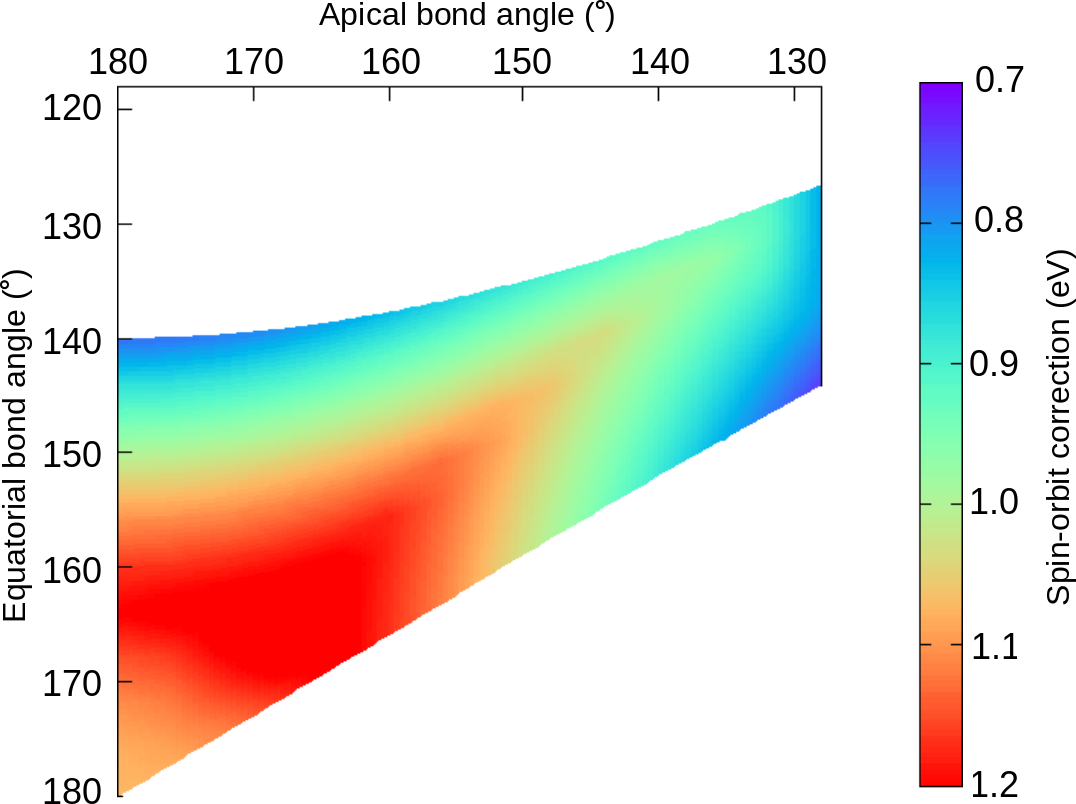}
  \end{center}
  \small
  {\bf Supplementary Figure 1\hspace{0.2cm}
  Two-dimensional map of the spin-orbit coupling correction to the scalar-relativistic DFT/LDA
  band gap.} The calculations were performed for the Platonic model of PbI$_3$-based perovskites 
  as a function of the apical and equatorial metal-halide-metal bond angles. Calculation details 
  are provided in the Methods. The region beyond $\sim$165$^\circ$ on the bottom left corner should
  not be considered as reliable, since the relativistic DFT band gap closes in this range
  (this is only an artefact).\vspace{5cm}
  \end{figure}

  \begin{figure}[p]
  \begin{center}
  \includegraphics[width=1\textwidth]{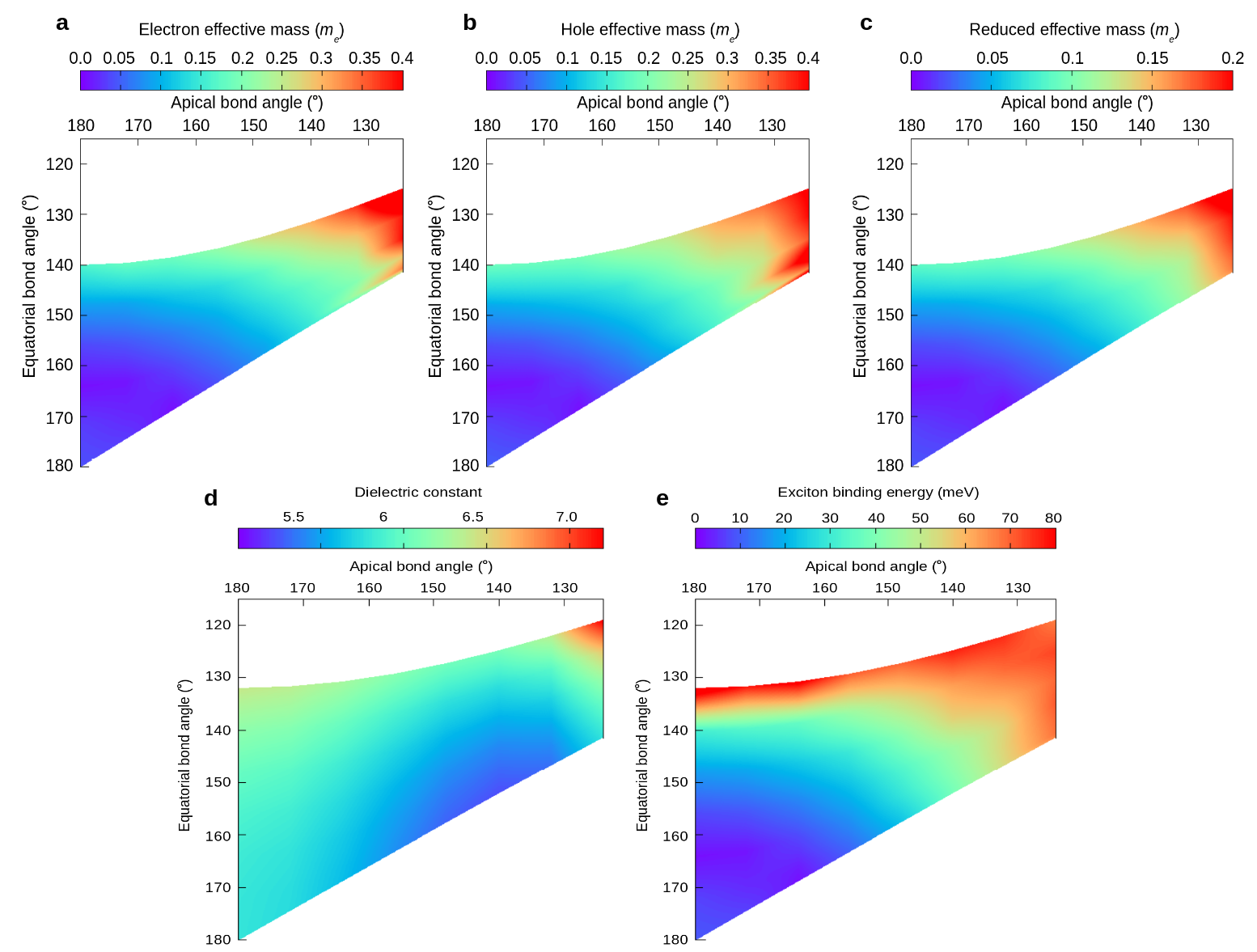}
  \end{center}
  \small
  {\bf Supplementary Figure 2\hspace{0.2cm}
  Calculated carrier effective masses, dielectric constant, and exciton binding energy  
  for the Platonic model of PbI$_3$-based perovskites}. {\bf a}-{\bf c}
  Isotropic average of the effective mass tensor for holes ({\bf a}), electrons ({\bf b}),
  and electron-hole pairs ({\bf c}).
  {\bf d}. The calculated static (electronic) dielectric constant.
  {\bf e} Exciton binding energy within the Wannier model.
  Effective masses are only meaningful for angles smaller than $\sim$165$^\circ$, since
  in the bottom left corner of the map the calculated band gaps vanish.\vspace{5cm}
  \end{figure}

  \begin{figure}[p]
  \begin{center}
  \includegraphics[width=0.8\textwidth]{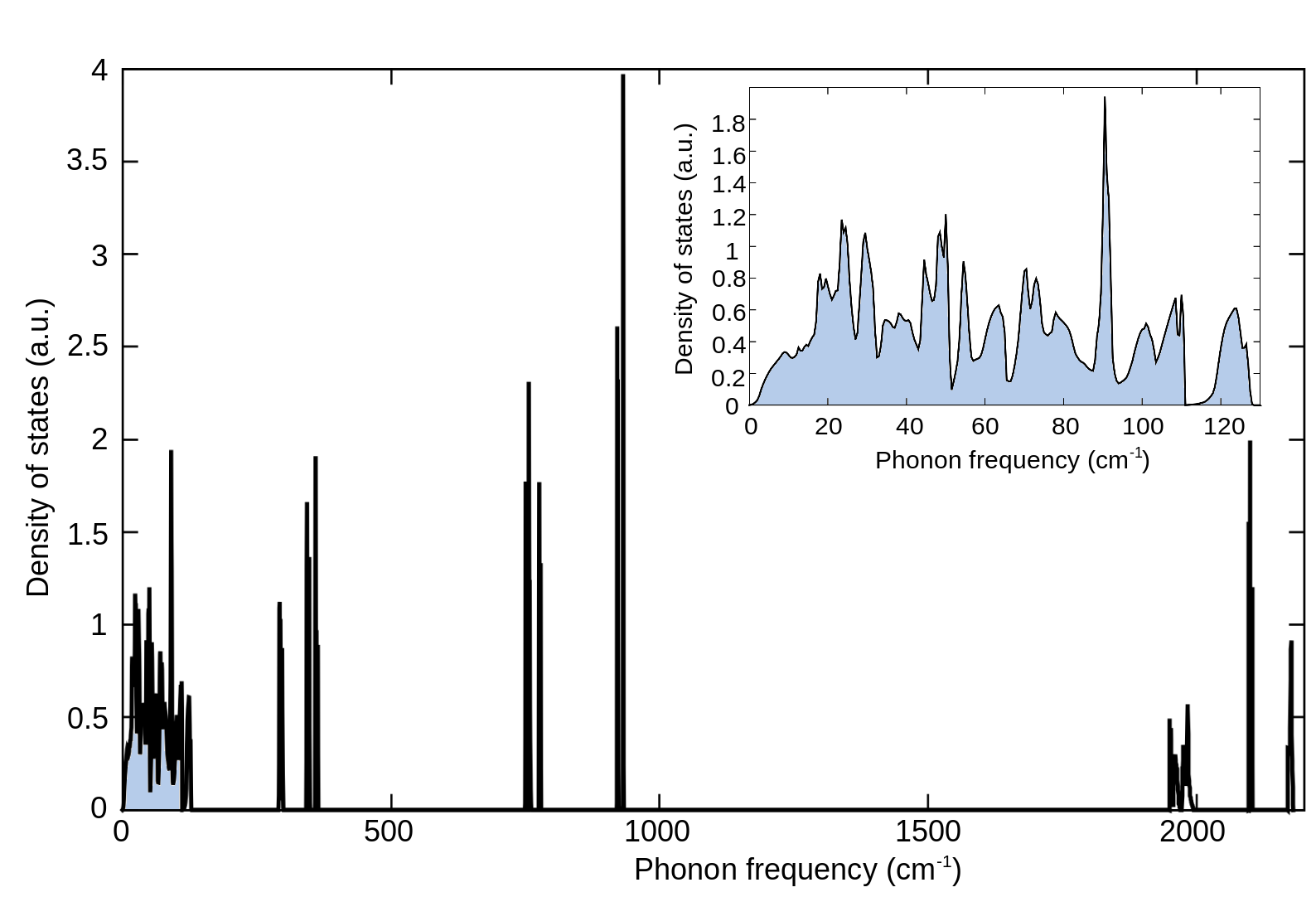}
  \end{center}
  \small
  {\bf Supplementary Figure 3\hspace{0.2cm} Calculated vibrational density of states of AsH$_4$PbI$_3$}.
  The calculations were performed using density functional perturbation theory \cite{Baroni},
  and a $2\times 2 \times 4$ Brillouin zone grid. 
  The vibrational density of states was then generated using the linear tetrahedron method 
  ~\cite{Blochl}. The inset shows a close up of the low-energy region of the spectrum.\vspace{1cm}
  \end{figure}

  \begin{figure}[p]
  \begin{center}
  \includegraphics[scale=0.27]{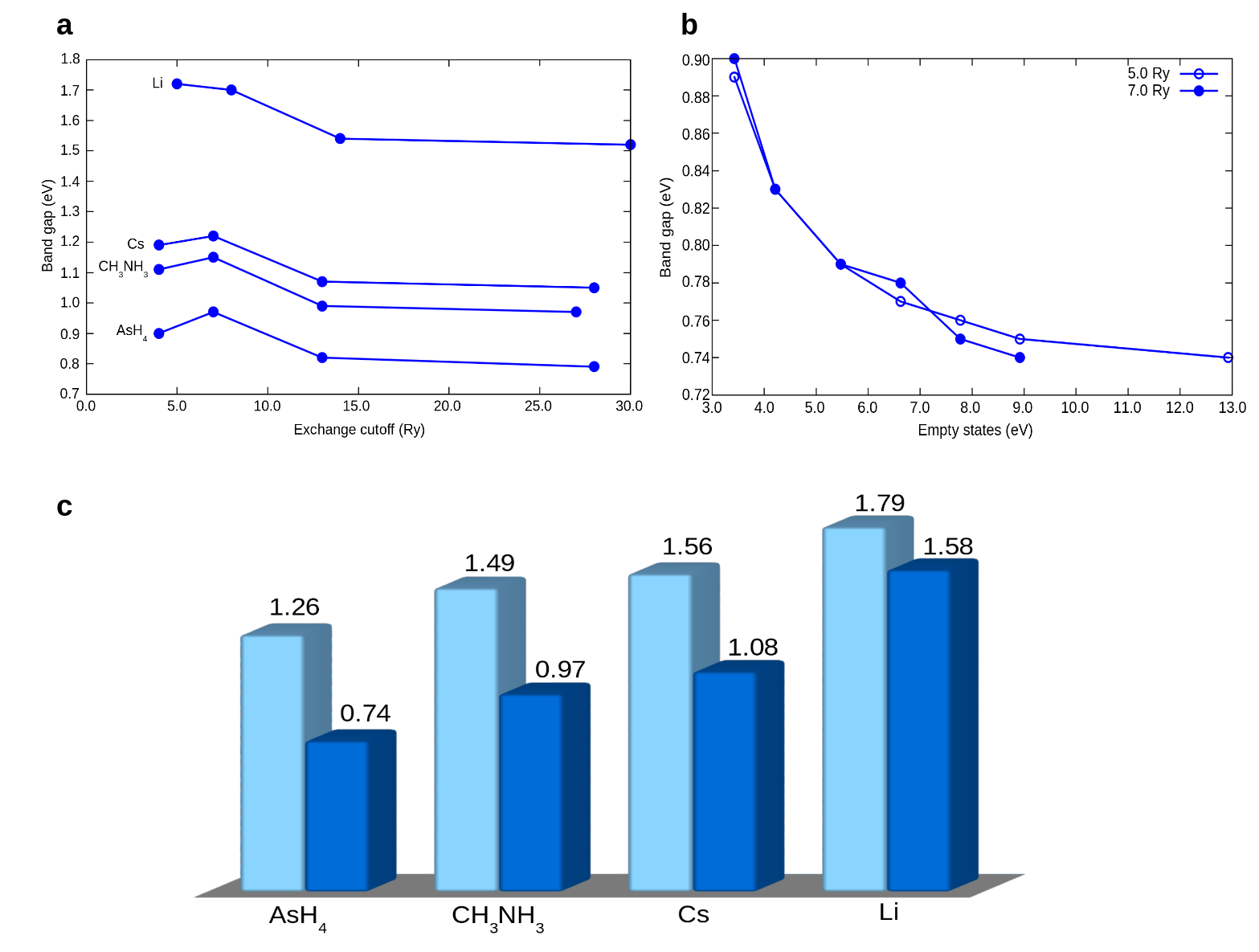}
  \end{center}
  \small
  {\bf Supplementary Figure 4\hspace{0.2cm} Quasiparticle band gaps}. {\bf a} Convergence of 
  the gap with respect to the planewaves kinetic energy cutoff
  of the exchange self-energy, for AsH$_4$PbI$_3$, CH$_3$NH$_3$PbI$_3$, CsPbI$_3$, and LiPbI$_3$.
  The calculations were performed using a $\Gamma$-point sampling, a cutoff of 5~Ry for the
  correlation self-energy, and 400 Kohn-Sham states.
  {\bf b} Convergence of the gap with respect to the expansion of the polarizability 
  over virtual Kohn-Sham states, for two different values of the correlation cutoff, 5~Ry (open
  circles) and 7~Ry (filled circles).
  The horizontal axis represents the eigenvalue of the highest 
  state included in such expansion, relative to the valence band top. The rightmost datapoint 
  corresponds to a calculation with 400 states. 
  In this case the calculations were performed for AsH$_4$PbI$_3$ using a 2$\times$2$\times$2 
  $\Gamma$-centered Brillouin zone mesh, and a cutoff of 20~Ry for the exchange part of the
  self-energy.
  \textbf{c} Comparison between scalar relativistic DFT/LDA band gaps (light rods) 
   and fully relativistic GW band gaps (dark rods). Fully-relativistic $GW$ calculations yield 
   lower band gaps but confirm the general trend. All band gaps are in units of eV.  
  \end{figure}

  \begin{table}
  \begin{center}
  \begin{tabular}{lcclc}
  \hline
  \hline
  \\[-8pt]
   Cation & Steric radius (\AA) & & Cation & Steric radius (\AA) \\
  \\[-8pt]
  \hline
  \\[-8pt]
  Li$^+$       & 0.67 & & AsH$_4^+$           & 2.14 \\
  Na$^+$       & 0.85 & & SbH$_4^+$           & 2.36 \\ 
  K$^+$        & 1.31 & & CH$_3$NH$_3^+$      & 2.03 \\ 
  Rb$^+$       & 1.50 & & CH$_3$PH$_3^+$      & 2.38 \\ 
  Cs$^+$       & 1.77 & & CH$_3$AsH$_3^+$     & 2.49 \\ 
  NH$_4^+$     & 1.56 & & CH$_3$SbH$_3^+$     & 2.72 \\ 
  NCl$_4^+$    & 2.72 & & CH$_2$N$_2$H$_4^+$  & 2.24 \\ 
  PH$_4^+$     & 2.04 & & C$_2$H$_6$NH$_2^+$  & 2.42 \\ 
  PF$_4^+$     & 2.14 & & C$_2$H$_6$PH$_2^+$  & 2.68 \\  
  \\[-10pt]
  \hline
  \hline
  \end{tabular}
  \end{center}
  \small
  {\bf Supplementary Table 1 \hspace{0.2cm} Steric radii of the cations}.
  We define the steric radius as the radius of the sphere which contains 95\% of
  the total electron density. The density is calculated using density functional theory, as
  described in the manuscript, for molecules in their singly ionized state. For comparison, 
  the ionic radii of Rb$^+$ and Cs$^+$ are 1.66~\AA~ and 1.81~\AA, respectively~\cite{Shannon}. 
  \end{table}

\newpage
\section*{Supplementary Notes}

{\bf Supplementary Note 1}

Using the effective masses and the dielectric constants thus obtained, we calculate the variation
of the exciton binding energy $E_{\rm b}$ shown in Supplementary Figure~2e. 
Our calculations are in line with recent measurements on CH$_3$NH$_3$PbI$_{3-x}$Cl$_x$
films~\cite{dInnocenzo2014}, which reported an exciton binding energy of 55 meV~\cite{dInnocenzo2014}.
Given the small variations of the dielectric
constant with the apical and equatorial bond angles, we note that the exciton binding energy
shows a similar dependence on the Pb-I-Pb bond angles as the effective mass and the band gap.
Therefore, structures with a lower degree of octahedral tilt should also exhibit lower binding energies.

{\bf Supplementary Note 2}

Supplementary Figures~4a and 4b show the convergence of the quasiparticle band gap
with the planewaves kinetic energy cutoff and with the number of virtual states,
for a few selected structures (AsH$_4$PbI$_3$, CH$_3$NH$_3$PbI$_3$, CsPbI$_3$, LiPbI$_3$).
In particular, Supplementary Figure~4a shows that the convergence tests for the four structures
considered here exhibit very similar behavior. This suggests that our predicted
trend for the band gaps should not be affected by numerical convergence.
We also tested the effect of Brillouin-zone sampling on the quasiparticle band gaps.
We find a difference of less that 0.1 eV when increasing the zone sampling from
1$\times$1$\times$1 to  4$\times$4$\times$3, irrespective of the planewaves
cutoff.

Our calculated quasiparticle correction does not fully compensate the red-shift due to
spin-orbit effects in any of the four test structures. As a result our relativistic $GW$
band gaps are smaller than the ones obtained within scalar-relativistic DFT/LDA 
(Supplementary Figure~4c).
This finding is in line with previous attempts at calculating the quasiparticle band gap 
of MA-PbI$_3$ and CsPbI$_3$ within a $G_0W_0$ calculation including spin-orbit 
coupling~\cite{Even2013, Brivio2013}. 

Moreover, our $GW$ band gaps of CH$_3$NH$_3$PbI$_3$ and CsPbI$_3$ underestimate experimental
optical gaps. This suggests that going beyond the $G_0W_0$ approximation may be necessary.
Here we speculate that the lack of self-consistency or the use of the plasmon pole model 
may be at the origin of this underestimation, as hinted in Ref.\cite{Brivio2013,Umari2014}.

While a fine-tuning of the theoretical methodology used in our calculations may lead to
closer agreement with experimental optical gaps for individual structures, in the present work 
we focus on band gap trends. Such trends are consistent throughout largely different levels of 
theory, from scalar-relativistic DFT, to fully-relativistic DFT, to relativistic many-body perturbation theory.

\section*{Supplementary methods}

\section*{\small Exciton binding energy}

In order to estimate the magnitude of excitonic effects on the band gaps
we calculate the exciton binding energies using a simple Wannier exciton model~\cite{Yu-book}. 
In the Wannier model the exciton binding energy is given by $E_{\rm b} = \mu^*/\varepsilon_0^2 E_{1s}$,
where $E_{1s}$ is the energy of the fundamental state of the hydrogen atom, $\mu^*$ is the effective
mass of electrons and holes (i.e.\ $1/\mu^* = 1/m_{\rm e} + 1/m_{\rm h}$ ), 
and $\varepsilon_0$ the static (electronic) dielectric constant of the solid. 
This model is very approximate but should provide a fair representation of the trends
in exciton binding energies in our compounds.

We calculated the effective mass tensors by approximating the second derivatives of the relativistic
Kohn-Sham eigenvalues via finite differences, and taking the isotropic average in each case.
Supplementary Figures~2a and 2b show the hole and electron effective masses thus obtained for the Platonic
perovskite model, as a function of equatorial and apical metal-halide-metal bond angles.
Our calculated effective masses are in line with previous calculations for
CH$_3$NH$_3$PbI$_3$ and CsPbI$_3$ \cite{Umari2014,Even2013, Even2014}. As expected
from the variation of the band gap with the bond angles, the electron and hole masses 
decrease when increasing the Pb-I-Pb bond angles.

For the dielectric constant we performed finite-electric field calculations using the Berry-phase 
technique~\cite{Vanderbilt2002, Umari2002}. In this case we adopted a $\Gamma$-centered 2$\times$2$\times$4 
Brillouin-zone mesh, with a finite electric field of 0.001 a.u.\ directed along the $c$-axis.
While in this case we did not not evaluate the isotropic average (purely in order to contain 
computational costs), Supplementary Figure~2d shows that this should not pose a problem since
$\varepsilon_0$ does not depend strongly on the bond angles in the range of interest ($\varepsilon_0=5.5-6.0$).

\section*{\small Band gaps from many-body perturbation theory}

In order confirm the band gap trend obtained within DFT/LDA 
we have performed $G_0W_0$ calculations using the \texttt{Yambo} code~\cite{Marini2009} for 
the following compounds: AsH$_4$PbI$_3$, CH$_3$NH$_3$PbI$_3$, CsPbI$_3$, LiPbI$_3$.
For these calculations we used LDA norm conserving fully relativistic pseudopotentials for Pb and I,
with semicore $5d$ states included for Pb. The Kohn-Sham eigenfunctions and eigenvalues were calculated
using a 100~Ry kinetic energy cutoff for all structures except LiPbI$_3$, for which we used a cutoff of 130~Ry.
The exchange and the correlation parts of the $GW$ self-energy were calculated using planewaves cutoffs
of 13~Ry and 5~Ry, respectively (this is at the limit of what we can afford given the large unit cells).
The screened Coulomb interaction $W$ was described by means of the Godby-Needs plasmon-pole
model~\cite{Godby}, and 400 bands were included in the sum over virtual states. The wavevector-dependence
of the screened Coulomb interaction was described by sampling the Brillouin zone using a
$\Gamma$-centered 2$\times$2$\times$2 grid.

%\newpage

\newpage

\section*{Supplementary References}

\end{document}